%% file: main.tex
\documentclass{IEEEcsmag}

\usepackage[colorlinks,urlcolor=blue,linkcolor=blue,citecolor=blue]{hyperref}

\usepackage{upmath}
\usepackage{caption}
\usepackage{subcaption}
\usepackage{balance}

\usepackage{mathtools}
\usepackage{color, colortbl}
\usepackage{ctable}
\definecolor{Gray}{gray}{0.9}
\usepackage{graphicx}
\usepackage{caption}
\usepackage{color, colortbl}
\usepackage{array}
\usepackage{booktabs} 
\usepackage{makecell}
\usepackage{cellspace}
\usepackage{tabularx}
\usepackage{multirow}
\usepackage{amsmath}

\usepackage{xurl}

\usepackage{xcolor}

\usepackage{breakcites}
\usepackage{multirow}

\jvol{XX}
\jnum{XX}
\paper{8}

\begin{document}


\title{Good Intentions, Bad Inventions: How Employees Judge Pervasive Technologies in the Workplace}

\author{\ Marios Constantinides}
\affil{\ Nokia Bell Labs, Cambridge (UK)}
\editor{Corresponding author: Marios Constantinides, marios.constantinides@nokia-bell-labs.com}

\author{\ Daniele Quercia}
\affil{\ Nokia Bell Labs, Cambridge (UK)}


\begin{abstract}
Pervasive technologies combined with powerful AI have been recently introduced to enhance work productivity. Yet, some of these technologies are judged to be invasive. To identify which ones, we should understand how employees tend to judge these technologies. We considered 16 technologies that track productivity, and conducted a study in which 131 crowd-workers judged these scenarios. We found that a technology was judged to be right depending on the following three aspects of increasing importance. That is, whether the technology: \emph{1)} was currently supported by existing tools; \emph{2)} did not interfere with work or was fit for purpose; and \emph{3)} did not cause any harm or did not infringe on any individual rights. Ubicomp research currently focuses on how to design better technologies by making them more accurate, or by increasingly blending them into the background. It might be time to design better ubiquitous technologies by unpacking AI ethics as well.
\end{abstract}

\maketitle


\input{sections/1_Introduction}
\input{sections/2_RelatedWork}
\input{sections/3_OnlineStudy}
\input{sections/4_Analysis}
\input{sections/5_Results}
\input{sections/6_Discussion}

\bibliographystyle{abbrv} 

\bibliography{bibliography}  

\begin{IEEEbiography}{Marios Constantinides}{\,}is currently a Senior Research Scientist at Nokia Bell Labs Cambridge (UK). He works in the areas of human-computer interaction and ubiquitous computing. Contact him at marios.constantinides@nokia-bell-labs.com.
\end{IEEEbiography}

\begin{IEEEbiography}{Daniele Quercia}{\,}is currently the Department Head at Nokia Bell Labs in Cambridge (UK) and Professor of Urban Informatics at King's College London. He works in the areas of computational social science and urban informatics. Contact him at quercia@cantab.net.
\end{IEEEbiography}

\end{document}

%% file: sections/1_Introduction.tex
\chapterinitial{The introduction} 
New pervasive technologies in the workplace have been introduced to enhance productivity (e.g., a tool that provides an aggregated productivity score based on, for example, email use on the move, network connectivity, and exchanged content). Yet, some of them are judged to be invasive so much so that they make it hard to build a culture of trust at work, and often impacting workers' productivity and well-being in negative ways~\cite{alge2013workplace}. While these technologies hold the promise of enabling employees to be productive, report after report has highlighted the outcries of AI-based tools being biased and unfair, and lacking transparency and accountability~\cite{buolamwini2018gender}. Systems are now being used to analyze footage from security cameras in workplace to detect, for example, when employees are not complying with social distancing rules\footnote{\url{https://www.ft.com/content/58bdc9cd-18cc-44f6-bc9b-8ca4ac598fc8}}; while there is a handful of good intentions behind such a technology (e.g., ensuring safe return to the office after the COVID-19 pandemic), the very same technology could be used to surveil employees' movements, or time away from desk. Companies now hold protected intellectual properties on technologies that use ultrasonic sound pulses to detect worker's location and monitor their interactions with inventory bins in factories.\footnote{Wrist band haptic feedback system: \url{https://patents.google.com/patent/WO2017172347A1/}}

As we move towards a future likely ruled by big data and powerful AI algorithms, important questions arise relating to the psychological impacts of surveillance, data governance, and compliance with ethical and moral concerns (\url{https://social-dynamics.net/responsibleai}). To make the first steps in answering such questions, we set out to understand how employees judge pervasive technologies in the workplace and, accordingly, determine how desirable technologies are supposed to behave both onsite and remotely. In so doing, we made two sets of contributions: First, we considered 16 pervasive technologies that track workplace productivity based on a variety of inputs, and conducted a study in which 131 US-based crowd-workers judged these technologies along the 5 well-established moral dimensions of harm, fairness, loyalty, authority, and purity~\cite{haidt2007new}. We found that the judgments of a scenario were based on specific heuristics reflecting whether the scenario: was currently supported by existing technologies; interfered with current ways of working or was not fit for purpose; and was considered irresponsible by causing harm or infringing on individual rights. Second, we measured the moral dimensions associated with each scenario by asking crowd-workers to associate words reflecting the five moral dimensions with it. We found that morally right technologies were those that track productivity based on task completion, work emails, and audio and textual conversations during meetings, whereas morally wrong technologies were those that involved some kind of body-tracking such as tracking physical movements and facial expressions.

%% file: sections/2_RelatedWork.tex
\section{RELATED WORK}
\label{sec:related_work}

On a pragmatic level, organizations adopted  ``surveillance'' tools mainly to ensure security and boost productivity~\cite{ball2010workplace}. In a fully remote work setting, organizations had to adopt new security protocols~\cite{mckinsey} due to the increased volume of online attacks,\footnote{\url{https://www.dbxuk.com/statistics/cyber-security-risks-wfh}} and they 
ensured productivity by tracking the efficient use of resources~\cite{ball2010workplace}.

However, well-meaning technologies could inadvertently be turned into surveillance tools. For example, a technology that produces an aggregated productivity score\footnote{\url{https://www.theguardian.com/technology/2020/nov/26/microsoft-productivity-score-feature-criticised-workplace-surveillance}} based on diverse inputs (e.g., email, network connectivity, and exchanged content) can be a double-edged sword. On the one hand, it may provide managers and senior leadership visibility into how well an organization is doing. On the other hand, it may well be turned into an evil tool that puts employees under constant surveillance and unnecessary psychological pressure.\footnote{\url{https://twitter.com/dhh/status/1331266225675137024}} More worryingly, one in two employees in the UK thinks that it is likely that they are being monitored at work~\cite{tucsurvey}, while more than two-thirds are concerned that workplace surveillance could be used in a discriminatory way, if left unregulated. Previous studies also found that employees are willing to be `monitored' but only when a company's motivations for doing so are transparently communicated~\cite{marchant2019best}. Technologies focused on workplace safety typically receive the highest acceptance rates~\cite{jacobs2019employee}, while technologies for unobtrusive and continuous stress detection receive the lowest,  with employees mainly raising concerns about tracking privacy-sensitive information~\cite{kallio2021unobtrusive}.

To take a more responsible approach in designing new technologies, researchers have recently explored which factors affect people's judgments of  these technologies.  In his book \emph{``How humans judge machines''}~\cite{hidalgo2021humans}, Cesar Hidalgo showed that people do not judge humans and machines equally, and that differences were the result of two principles. First, people judge humans by their intentions and machines by their outcomes (e.g., \emph{``in natural disasters like the tsunami, fire, or hurricane scenarios, there is evidence that humans are judged more positively when they try to save everyone and fail---a privilege that machines do not enjoy''~\cite{hidalgo2021humans}-p. 157)}. Second, people assign extreme intentions to humans and narrow intentions to machines, and, surprisingly, they may excuse human actions more than machine actions in accidental scenarios
(e.g., \emph{``when a car accident is caused by either a falling tree or a person jumping in front of a car, people assign more intention to the machine than to the human behind the wheel''~\cite{hidalgo2021humans}-p. 130)}.

Previous work has mostly focused on scenarios typically involving aggression, physical, or psychological harm. Here, in the workplace context, we explore scenarios reflecting aspects tailored to the pervasive computing research agenda that transcend harm such as ease of adoption and technological intrusiveness. 

%% file: sections/3_OnlineStudy.tex
\section{ONLINE STUDY}
\label{sec:study}

\begin{table*}[t]
\centering
\caption{Summary of crowd-workers demographics. }
\begin{tabular}{|l l|} 
 \hline
 Type & Count\\  \hline
  Gender &  Male (66\%), Female (34\%)\\
  Ethnicity &  White (80\%), African-American (13\%), Asian (4\%) Hispanic (3\%)\\
  Years of employment & Less than 2 years (24\%), 2-5 years (53\%), 5+ years (23\%) \\
  Time working remotely &  Less than years (67\%), 2-5 years (20\%), 5+ years (5\%), Never (8\%)\\
  Industry sector & {\shortstack [l]{IT (40\%), Financials (21\%), Industrials (12\%), Energy (11\%), Health Care (6\%), Materials (4\%) \\  Consumer Staples (2\%), Communication Service (2\%), Consumer Discretionary (2\%)}} \\
  Role &  {\shortstack [l]{Manager (54\%), Software Engineer (17\%), Sales and Marketing (5\%), Accountant (5\%) \\ Office administrator (3\%), Human Resources (2\%) not specified (14\%)}}\\
 \hline
\end{tabular}
\label{tab:crowdworkers_demographics}
\end{table*}

\subsection{Scenarios Generation}
\label{sec:workplace_technologies}
The Electronic Frontier Foundation, a leading non-profit organization defending digital privacy and free speech has analyzed employee-monitoring software programs~\cite{eef_analysis}, and classified these programs based on five main aspects that are being tracked: (a) work time on computer (e.g., tracking inactivity) (b) log keystrokes (e.g., typing behavior, text messages being exchanged), (c) websites, apps, social media use, and emails, (d) screenshots to monitor task completion time, and (e) webcams monitoring facial expressions, body postures, or eye movements. Drawing from this analysis, we devised a set of 16 AI-based workplace technologies (Table~\ref{tab:technologies}). As a result of rapid technological advancements, this list might not be exhaustive, but, as we shall see next (\S\nameref{sec:analysis}), our methodology could be used on newly introduced technologies as it is a generalizable way of identifying how individuals tend to make their moral judgments.

\begin{table*}[t]
\centering
\caption{Sixteen tracking technologies that were judged along five well-established moral dimensions: harm, fairness, loyalty, authority, and purity.}
\begin{tabular}{|l l|} 
 \hline
 Tracking technology & Example\\  \hline
  
 (1) body postures outside meetings & An earbud device tracking body postures through inertial measurement unit (IMU) data \\
 (2) body postures during meetings & An earbud device tracking body postures through IMU data \\
 (3) facial expressions outside meetings & A camera recording and analyzing an employee's face outside a meeting\\
 (4) facial expressions during meetings & A camera recording and analyzing an employee's face during a meeting\\
 (5) eye movements outside meetings & A camera or smart-glasses recording and analyzing an employee's face\\
 (6) eye movements during meetings & A camera or smart-glasses recording and analyzing an employee's face\\
 (7) video streams during meetings & A camera recording an employee's face and body\\
 (8) audio conversations during meetings & A microphone recording a meeting's conversation\\
 (9) text exchanges during meetings & A software tracking textual conversations during a meeting\\
 (10) physical movements & A camera- or IMU-based tracking device that infers physical movements\\
 (11) work emails & A software accessing and analyzing emails\\
 (12) applications used & A software tracking applications in an employee's workstation \\
 (13) websites visited & A software tracking sites an employee visited\\
 (14) social media use & A software recording an employee's social media activity \\
 (15) tasks completion & A software tracking a to-do list where one marks the completed tasks  \\
 (16) typing behavior & A keylogger software installed in an employee's workstation \\
 \hline
\end{tabular}

\label{tab:technologies}
\end{table*}

Having the 16 technologies at hand, we created scenarios involving their use onsite and remotely. Scenarios are short stories that describe an action that can have a positive or negative moral outcome~\cite{hidalgo2021humans}. Here, an action is defined as a technology that tracks productivity through certain types of data. For example, the scenario for \emph{tracking productivity through social media use (Technology 14 in Table~\ref{tab:technologies})} when working remotely reads as: 
\emph{All employees are working remotely and, as a new policy, their company is using the latest technologies to keep track of their social media use to monitor productivity.} Having 16 technologies and 2 work modes (i.e., onsite or remotely), we ended up with 32 scenarios.

\subsection{Procedure}
\label{sec:procedure}
For each scenario, we used a set of questions probing people's attitudes toward a technology. We captured these attitudes through three questions (facets) concerning whether a technology is: hard to adopt, intrusive, and harmful. These facets originate from experiments conducted to understand people's attitudes toward AI more generally~\cite{hidalgo2021humans} (p. 27). For each scenario, we asked three questions, answered on a Likert-scale:

\begin{enumerate}
    \item Was the technology hard to adopt? \\ (1: extremely unlikely; 7: extremely likely)
    \item Was the technology intrusive? \\ (1: extremely unobtrusive; 7: extremely intrusive)
    \item Was the technology  harmful? \\ (1: extremely harmless; 7: extremely harmful)
\end{enumerate}

After responding to these questions, crowd-workers were asked to choose words associated with five moral dimensions that best describe the scenario. In general, morality speaks to what is judged to be ``right'' or ``wrong'', ``good'' or ``bad''. Moral psychologists identified a set of five dimensions that influence individuals' judgments~\cite{haidt2007new}: \emph{harm} (which can be both physical or psychological), \emph{fairness} (which is typically about biases), \emph{loyalty} (which ranges from supporting a group to betraying a country), \emph{authority} (which involves disrespecting elders or superiors, or breaking rules), and \emph{purity} (which involves concepts as varied as the sanctity of religion or personal hygiene).

Each dimension included two positive and two negative words~\cite{hidalgo2021humans} (p. 28). The dimension \emph{harm} included the words `harmful (-)', `violent (-)', `caring (+)', `protective (+)'; the dimension \emph{fairness} included the words `unjust (-)', `discriminatory (-)', `fair (+)', `impartial (+)'; the dimension  \emph{loyalty} included the words `disloyal (-)', `traitor (-)', `devoted (+)', `loyal (+)'; the dimension \emph{authority} included the words `disobedient (-)', `defiant (-)', `lawful (+)', `respectful (+)'; the dimension \emph{purity} included the words `indecent (-)', `obscene (-)', `decent (+)', `virtuous (+)'. The (-) and (+) signs indicate whether a word has a negative or positive connotation. In the work environment, some of these terms (e.g., violent, traitor) might not apply, and, as such, we studied all the words aggregated by moral dimension rather than studying them individually. Finally, to place our results into context, we asked crowd-workers to report their basic demographic information  (e.g., gender, ethnicity, industry sector, number of years of employment).

\subsection{Participants and Recruitment}
We administered the 32 scenarios through Amazon Mechanical Turk (AMT), which is a popular crowd-sourcing platform for conducting social experiments. We only recruited highly reputable AMT workers by targeting workers with 95\% HIT (Human Intelligence Task) approval rate and at least 100 approved HITs. We applied quality checks using an attention question, which took the form of ``Without speculating on possible advances in science, how likely are you to live to 500 years old?'' An attention question is a standard proactive measure to ensure data integrity~\cite{peer2014reputation}, which helps to detect and discard responses generated by inattentive respondents. To this end, we rejected those who chose any option other than the less likely, leaving us with a total of 131 crowd-workers with eligible answers. To ensure that crowd-workers had a common understanding of these technologies, we provided examples of how each technology could work in tracking employees. The scenarios were randomized, ensuring no ordering effect would bias the responses. The task completion time was, on average, 12 minutes, and each crowd-worker received 1\$ as a compensation. 

In our sample, crowd-workers were from the U.S (whose statistics are summarized in Table~\ref{tab:crowdworkers_demographics}). In total, we received responses from 87 male and 44 female with diverse ethnic backgrounds; White (80\%), African-American (13\%), Asian (4\%), and Hispanic (3\%). These crowd-workers also come from diverse work backgrounds, ranging from Information Technology (40\%) to Industrials (12\%) to Communications Services (2\%), and held different roles in their companies such as managerial positions (54\%), software engineers (17\%), among others.

\subsection{Ethical Considerations}
The study was approved by Nokia Bell Labs, and the study protocol stated that the collected data will be analyzed for research purposes only. In accordance to GDPR, no researcher involved in the study could have tracked the identities of the crowd-workers (the AMT platform also uses unique identifiers that do not disclose the real identity of the worker), and all anonymous responses were analyzed at an aggregated level. 

%% file: sections/4_Analysis.tex
\section{ANALYSIS}
\label{sec:analysis}

For each technology $tech$ and question $i$, we computed the average rating when $tech$ is deployed onsite, and the average rating when it is deployed remotely:

\begin{align*}
& \textit{$onsite(tech)_{i}$} = \textit{$tech$'s average onsite rating}, \nonumber \\
& \textit{$remote(tech)_{i}$} = \textit{$tech$'s average remote rating}, \nonumber 
\end{align*}

To then ease comparability, we $z$-scored these two values:
\begin{align*}
& z_{onsite}(tech)_i  =  \frac{onsite(tech)_i - \mu_{onsite}(tech)_i}{\sigma_{onsite}(tech)_i},  \nonumber \\
& z_{remote}(tech)_i  =  \frac{remote(tech)_i - \mu_{remote}(tech)_i}{\sigma_{remote}(tech)_i} \nonumber
\end{align*}
where $i$ is one of the three questions (hard to adopt, intrusive, harmful), $\mu_{onsite}(tech)$ and $\sigma_{onsite}(tech)$ are the average and standard deviation of the ratings for all technologies deployed onsite, and $\mu_{remote}(tech)$ and $\sigma_{remote}(tech)$ are the average and standard deviation of the ratings for all technologies deployed remotely.

%% file: sections/5_Results.tex
\section{RESULTS}
\label{sec:results}

\begin{figure*}
    \centering
     \includegraphics[width=0.93\textwidth]{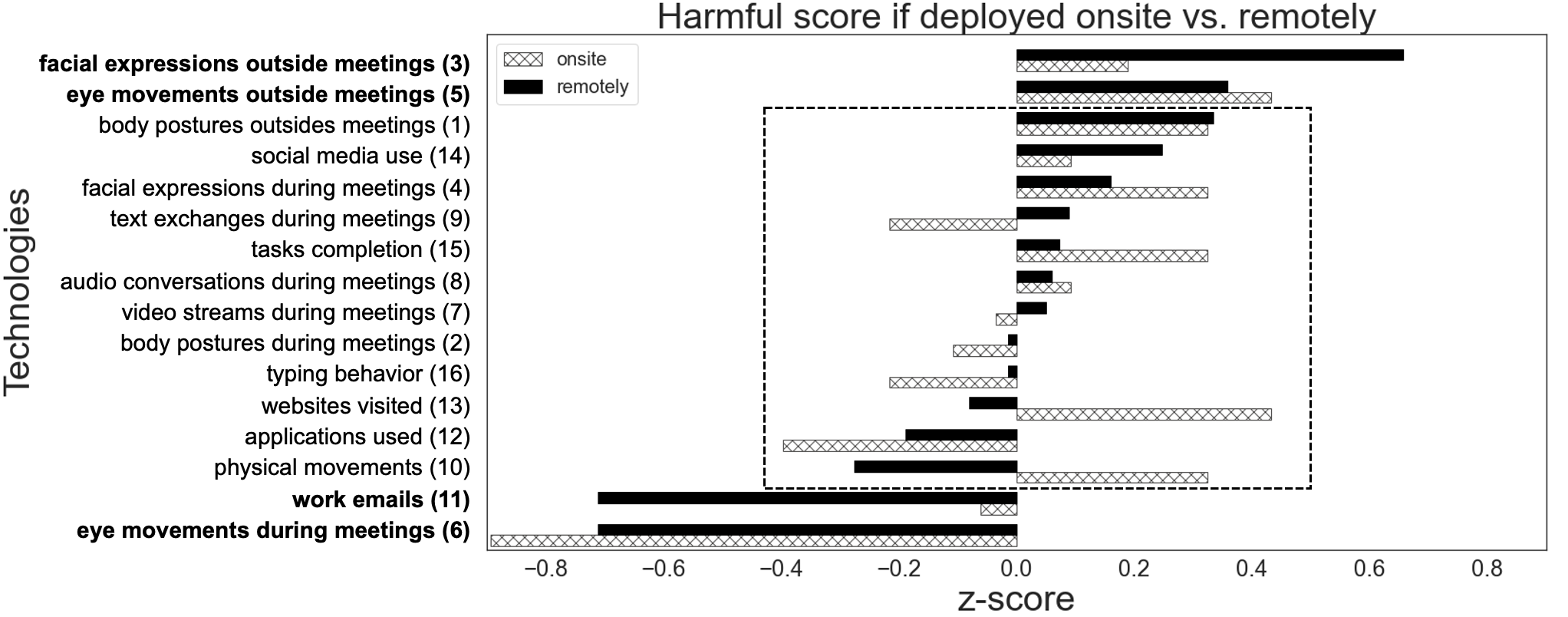}
    \caption{Four groups of technologies emerge: \emph{i)} technologies considered harmful (harmless) both onsite and remotely (inside the dashed box), \emph{ii)} technologies considered harmful (harmless) onsite but not remotely, or vice-versa (inside the dashed box), \emph{iii)} technologies considered harmful remotely (top bars outside the dashed box), and \emph{iv)} technologies considered harmless remotely (bottom bars outside the dashed box).  
    }
    \label{fig:harmful_quadrant}
\end{figure*}

\begin{table}[t]
\centering
\caption{Conditional probabilities of $p(row | column)$.}
\begin{tabular}{|l l l l|} 
 \hline
  & hard to Adopt & Intrusive  & Harmful\\  \hline
  hard to Adopt & 1 & 0.25 & 0.43 \\
  Intrusive & 0.2 & 1  & 0 \\
  Harmful & 0.6 & 0  & 1\\
 \hline
\end{tabular}
\label{tab:probs}
\end{table}

Unacceptable scenarios - those that were judged to be hard to adopt, intrusive, and harmful - include \emph{tracking physical movements}, especially \emph{onsite}. This scenario was indeed considered to: be not fully supported by current technologies in use (hard to adopt); interfere with work (intrusive); and infringe on one's freedom of movement (harmful).

As for all the scenarios, we tested how each of them was judged along multiple dimensions. To that end, we computed the conditional probability of a scenario that was judged, say, hard to Adopt to be also judged Harmful. This probability is $p(\textrm{Harmful} | \textrm{hard to Adopt}) $, and is equal to 0.6 (Table~\ref{tab:probs}), meaning that if a scenario is hard to adopt is also likely to be considered harmful, but not always, as we shall see next. More generally, the conditional probabilities are computed as:
$$
p(i | j) = \frac{\# \textrm{cases that are i and j}}{\# \textrm{cases that are  j}}$$

From these conditional probabilities, we can see that 20\% of the technologies that are hard to adopt are also considered intrusive, and 60\% are also considered harmful; 25\% of the technologies that are intrusive are considered hard to adopt; finally 43\% of the harmful technologies are considered hard to adopt.

By qualitatively analyzing the ways our participants motivated their judgments, we found that these judgments followed three main heuristics (i.e.,  mental shortcuts used to assess the scenarios quickly and efficiently):

\begin{itemize}

\item[] \textbf{Viability}. The first heuristic we identified was \textbf{whether the scenario can be easily built from existing technologies in a satisfactory manner}. This judgment criterion is associated with the moral dimension of fairness (any prototype of a technology that is hard to build would inevitably fall short and would be ridden with inaccuracies and biases). For example, in an online meeting, accurately \emph{tracking facial expressions} is technically easy to do using webcams. By contrast, \emph{tracking body postures} is still a hard problem because it requires a combination of wearable sensors such as multiple gyroscopes (e.g., a couple on the earphones~\cite{choi2021kairos, kawsar2018earables}, and the other on a smart watch), which ends up producing spurious classifications of body postures. \\

\item[] \textbf{Non-intrusiveness}. Another heuristic was \textbf{whether the scenario did not interfere with work or, more generally, was fit for purpose.} This judgment criterion was associated with the two moral dimensions of authority (when the technology is invasive and authoritarian) and loyalty (when the technology disrespects one's way of working and, as often mentioned by our respondents, it has been misused). For example, \emph{tracking eye movements in online meetings}, despite being possible, was considered to be unfit for productivity tracking  and be ``on the way'' of getting the job done. By contrast, \emph{tracking text messages in collaboration tools} such as Slack was considered to not interfere with work (unobtrusive) and fit for the purpose of tracking productivity (not misused). \\

\item[]  \textbf{Responsibility}. The final heuristic was \textbf{whether the scenario was considered responsible in that it did not cause any harm, or infringed on any individual rights.} This was associated with the two moral dimensions of harm (when the technology has negative effects on individuals) and purity (when the technology is seen to disrespect one's beliefs). For example, \emph{tracking audio conversations in online meetings} was considered to be possible (viable), and fit for purpose, yet it was considered to be harmful, as it entailed tracking not only whether a meeting took place but also its content.

\end{itemize}

\begin{figure*}
    \centering
     \includegraphics[width=0.92\textwidth]{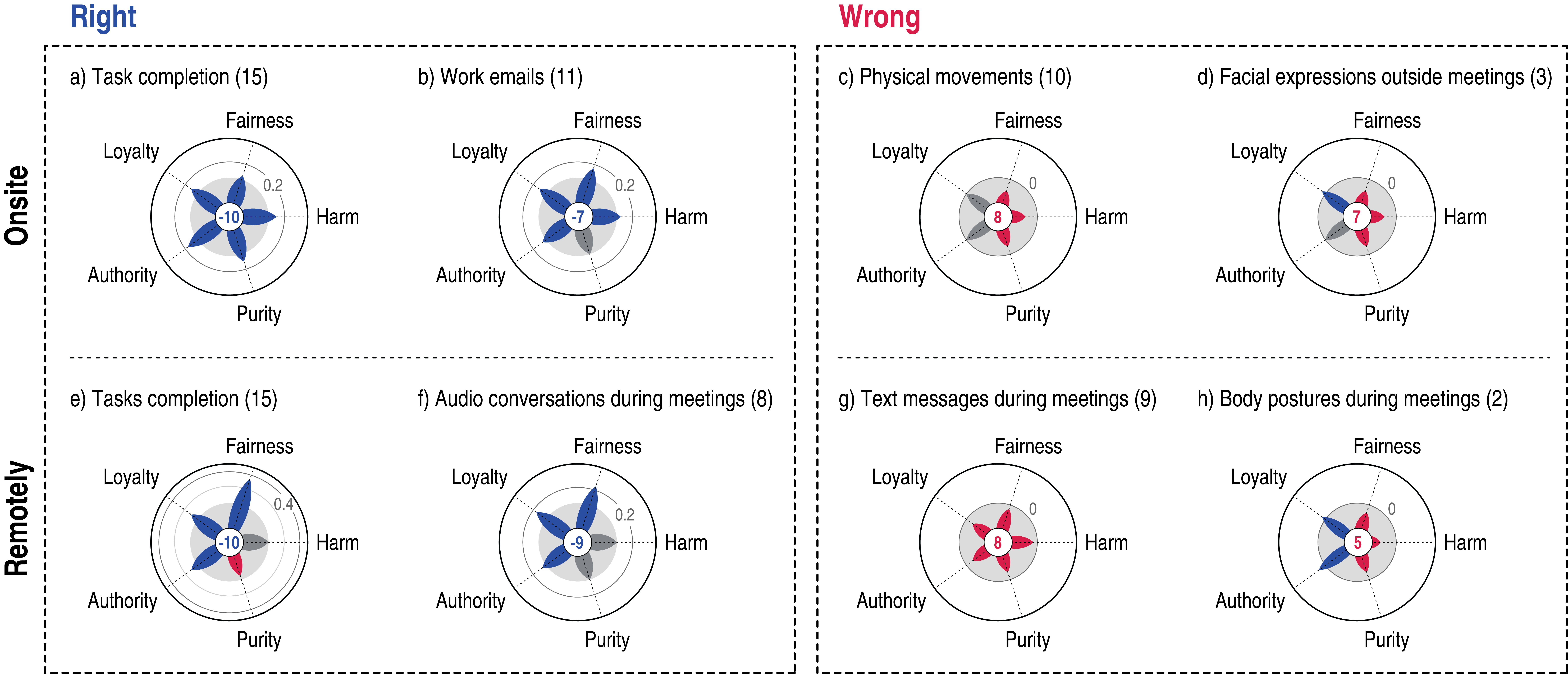}
    \caption{\textbf{a-b:} Top 2 most ``morally right'' technologies when applied onsite, and \textbf{c-d:} top 2 most ``morally wrong''  technologies when applied onsite. \textbf{e-f:} Top 2 most ``morally right''  technologies when applied remotely, and \textbf{g-h:} top 2 most ``morally wrong'' technologies when applied remotely. Each technology is marked with its number in Table~\ref{tab:technologies} in parenthesis. The wrongness score (showed at the center of each radar plot) was computed by aggregating the negative and the positive words of the individual five moral dimensions (\S\nameref{sec:procedure}), and captures how morally wrong the technology was considered on average to be. Dark blue denotes more positive words associated with a dimension, red denotes more negative words, and gray blue denotes equal amount of positive and negative words.}
    \label{fig:radar_plots}
\end{figure*}

\subsection{Onsite vs. Remotely: Clustering how technologies are judged}
Figure~\ref{fig:harmful_quadrant} shows the harmful score when a technology is deployed onsite versus when it is deployed remotely. Four groups of technologies emerge: \emph{a)} technologies considered harmful (harmless) both onsite and remotely, \emph{b)} technologies considered harmful (harmless) onsite but not remotely, or vice-versa, \emph{c)} technologies considered harmful remotely, and \emph{d)} technologies considered harmless remotely. As for the first two groups of technologies, for example, \emph{tracking audio conversations during meetings} led to same judgments irrespective the deployment setting, while, \emph{tracking text exchanges during meetings} led to opposite judgments. As for the third group, \emph{tracking eye movements} and \emph{facial expressions outside meetings} were considered more harmful when deployed remotely, not least because remote work typically happens in a private space~\cite{nippert2010islands} such as one's home~\cite{rudnicka2020eworklife}, and, as a result, the use of tracking devices should be limited to specific work-related activities and should ideally not go beyond them. As for the fourth group, \emph{tracking work emails} and \emph{tracking eye movements during meetings} were considered less harmful when deployed remotely. As it is harder to measure productivity in remote settings, tracking work emails was considered a reasonable proxy for attention levels and a compromise to accept\footnote{\url{https://www.gartner.com/smarterwithgartner/9-future-of-work-trends-post-covid-19}}\footnote{\url{https://www.computerworld.com/article/3586616/the-new-normal-when-work-from-home-means-the-boss-is-watching.html}}. Also, \emph{tracking eye movements during meetings} was seen as a proxy for body cues (e.g., facial expressions), which could reflect attention levels and often go unnoticed in virtual meetings~\cite{choi2021kairos}.

\subsection{Words associated with moral dimensions}
Next, we looked into how crowd-workers associated our technologies with words related to the five moral dimensions of \emph{harm, fairness, loyalty, authority, and purity}. We computed the fraction of times a word associated with each moral dimension was chosen (Figure~\ref{fig:radar_plots}). For example, \emph{tracking task completion onsite} (technology 15 in Table~\ref{tab:technologies}) was associated with fairness, loyalty, authority, purity, and lack of harm. The very same technology though when applied remotely was again associated with fairness, loyalty, authority, but also with harm and lack of purity.

By aggregating the negative and the positive words of the five moral dimensions (as per signs in \S\nameref{sec:procedure}), we computed a scenario's `wrongness'. This score captures how morally wrong or right a technology is. Technologies that were considered to be morally right (blue in Figure~\ref{fig:radar_plots}) were associated with positive words (e.g., fair, impartial), while those considered to be morally wrong (red in Figure~\ref{fig:radar_plots}) were associated with negative words (e.g., unjust, discriminatory). We found that morally right technologies with negative values of wrongness in Figures~\ref{fig:radar_plots}a-b, and Figures~\ref{fig:radar_plots}e-f were those that track productivity based on task completion, work emails, and audio and textual conversations during meetings, whereas morally wrong technologies (Figure~\ref{fig:radar_plots}c-d, and g-h) were those that involved some kind of body-tracking such as tracking physical movements and facial expressions. 

%% file: sections/6_Discussion.tex
\section{DISCUSSION AND CONCLUSION}
\label{sec:discussion}

\subsection{Discussion of Results}
We know surprisingly little about how people perceive pervasive technologies in the workplace. Yet we need to know more to inform the design of such technologies. This appears of crucial importance, not least because of the two polar opposite views animating today's debate over technology. On one hand, informed by a widespread algorithmic aversion, we risk rejecting technologies that could make our lives better. On the other hand, informed by technological optimism, we may adopt technologies that could have detrimental impacts. To avoid rejecting good technologies and designing bad ones, we should unpack AI ethics in Pervasive Computing. 

The heuristics we have found offer a guide on how technologies are likely to be morally judged. Having a technology that is easy to implement and does not interfere with work is not necessarily a technology that should be deployed. \emph{Tracking facial expressions} (even beyond the nefarious uses---of dubious effectiveness---of inferring political orientation or sexual preferences from faces~\cite{wang2018deep}) is possible and could be done in seamless ways (e.g., with existing off-the-shelf cameras), yet it would be still considered harmful and unethical. \emph{Tracking eye movements}, \emph{task completion}, or \emph{typing behavior} was considered a proxy for focus (harmless) yet intrusive as it would ``get in the way.'' \emph{Tracking social media use in remote work} was considered not only intrusive but also harmful, as it infringes on privacy rights. 

Finally, regardless of the work context (on-site \emph{vs.} remote work), most scenarios are either harmless (e.g., \emph{tracking application usage} was considered to be a proxy for  focus on work) or harmful (e.g., \emph{tracking physical movements, body posture, or facial expressions} was not). Yet, other scenarios were context dependent (Figure~\ref{fig:harmful_quadrant}). \emph{Tracking text messages during meetings} was considered less harmful (more fit-for-purpose) onsite than remotely. Text messages in onsite settings were considered ``fair game'' as they could reflect a meeting's productivity, while those in a remote setting were usually used beyond the meeting's purpose (e.g., used to catch up with colleagues), making them a poorer proxy for meeting productivity. Again, the heuristic used remained the same: whether a technology (e.g., inferring productivity from text messages) was fit for purpose. The only difference was that the same technology was fit for purpose in one context (e.g., in the constrained setting of a meeting) but not in the other (e.g., in the wider context of remote work). 

\subsection{Limitations and Future Work}
This work has three main limitations that call for future research efforts. The first limitation concerns the generalizability of our findings. While our sample demographics were fairly distributed across industry sectors and ethnic backgrounds, most of the crowd-workers were based in the US. Our findings hold for this specific cohort. The second limitation concerns the negative connotation of the three questions being asked in the crowd-sourcing experiment, which might have biased the responses. Even in such a case, the results would still make sense in a comparative way as the responses would be systematically biased across all scenarios. The third limitation concerns the pervasive technologies under study. Given the rapid technological advancements, at the time of writing, the 16 technologies in Table~\ref{tab:technologies} were considered of likely adoption. Future studies could replicate our methodology to larger and diverse cohorts, in specific corporate contexts or geographical units, and to 
emerging technologies such as AR headsets and EMG body-tracking devices.

\subsection{Theoretical and Practical Implications}
From a theoretical standpoint, it contributes to the ongoing debate of ethical and fair use of AI~\cite{arrieta2020explainable}---the emergent field of Responsible AI\footnote{\url{https://www.bell-labs.com/research-innovation/responsible-ai}}. As we showed, one needs to consider whether a technology is irresponsible in the first place well before its design. While tracking facial expressions is supported from current technology (viable) and can be done in seamless ways (unobtrusive), it was yet considered to be irresponsible (causing harm) in the office context. This translates into saying that companies should be more thoughtful about the ways they manage their workforce, and not deploying tools just because the technology allows them to. To determine whether a technology is irresponsible is a complex matter though, not least because it entails ethical concepts that are hard to define. That is why new approaches helping AI developers and industry leaders think about multi-faceted Responsible AI concepts should be developed in the near future.
From a practical standpoint, the Ubicomp community currently focuses on how to design better technologies by blending them into the background:  monitoring movements in a building through wifi signals~\cite{adib2013see}, for example, remain hidden. The problem is that, by blending technologies into the background, individuals are unaware of them and, as a result, their ethical concerns are often overlooked. That is why the Ubicomp community's aspiration of blending technologies needs to go hand-in-hand with the need of unpacking AI ethics.